\documentclass[sn-mathphys,Numbered]{sn-jnl}

\usepackage{graphicx}%
\usepackage{multirow}%
\usepackage{amsmath,amssymb,amsfonts}%
\usepackage{amsthm}%
\usepackage{mathrsfs}%
\usepackage[title]{appendix}%
\usepackage{xcolor}%
\usepackage{textcomp}%
\usepackage{manyfoot}%
\usepackage{booktabs}%
\usepackage{algorithm}%
\usepackage{algorithmicx}%
\usepackage{algpseudocode}%
\usepackage{listings}%

\theoremstyle{thmstyleone}%
\theoremstyle{thmstyletwo}%

\theoremstyle{thmstylethree}%

\begin{document}

\title[Entropy in black hole transformations]{Gravitational entropy in black hole transformations}


\author*[1]{\fnm{Alan Alejandro} \sur{Panuco Liñan}}\email{alejandro.panucolnn@uanl.edu.mx}

\author[2]{\fnm{Daniela} \sur{P\'erez}}\email{danielaperez@iar.unlp.edu.ar}

\author[1]{\fnm{Carlos} \sur{Luna}}\email{carlos.lunacd@uanl.edu.mx}

\author[2,3]{\fnm{Gustavo E.} \sur{Romero}}\email{romero@iar.unlp.edu.ar}


\affil*[1]{\orgname{Universidad Aut\'onoma de Nuevo Le\'on (UANL)},  \city{San Nicol\'as de los Garza}, \postcode{66455}, \state{Nuevo Le\'on}, \country{Mexico}}

\affil[2]{\orgname{Instituto Argentino de Radioastronom\'ia (IAR, CONICET/CIC/UNLP)},  \city{Villa Elisa}, \postcode{C.C.5, (1894)}, \state{Buenos Aires}, \country{Argentina}}

\affil[3]{\orgdiv{Facultad de Ciencias Astron\'omicas y Geofísicas de La Plata}, \orgname{UNLP}, \city{La Plata}, \postcode{1900}, \state{Buenos Aires}, \country{Argentina}}



\abstract{There are several reasons to support the idea that entropy might be associated to gravity itself. In the absence of a quantum theory of gravity, classical estimators for the gravitational entropy have been proposed. Any viable description of the gravitational entropy should reproduce the Hawking-Bekenstein entropy at the event horizon of black holes. Furthermore, in any black hole transformation, these estimators must satisfy the second law of black hole thermodynamics. In this work, we analyze whether two entropy estimators, one based on the Weyl tensor and the other on the Bel-Robinson tensor, satisfy the second law in the transformation process from a Schwarzschild to a Reissner-Nordström black hole by the absorption of a charge test particle. We also address the inverse process. We show that depending whether the process is reversible or not, both estimators fulfill the second law.}




\maketitle

\section{Introduction}

In the 1970s it was discovered that black holes have some properties similar to those of thermodynamic systems. One defining property of black holes is the presence of an event horizon. Some early work  \cite{Ch+70,Penrose1971,Hawking1971} showed that the area of the horizon, under a series of processes and transformations (Penrose process, black hole transformations), never decreases. Bekenstein was the first to propose that the area of the event horizon is proportional to the entropy of the black hole. He also formulated a generalized second law of black hole thermodynamics: in any physical process, the sum of the black hole entropy plus the entropy associated with ordinary matter never decreases \cite{Bekenstein1974a}. Bardeen, Carter, and Hawking extended this understanding by formulating four laws of black hole mechanics analogous to the four laws of thermodynamics, with the event horizon area and surface gravity corresponding to entropy and temperature of the system, respectively \cite{BA+73}. 

Classically, the analogy between the law of black hole mechanics and the laws of thermodynamics was only formal. However, if quantum effects are taken into account, as Zel'dovich and Starobinsky showed \cite{Zeldovich1971}, black holes should create and emit particles from background quantum fields.  This laid the groundwork for Hawking's formulation of thermal emission from black holes, known as Hawking radiation. This radiation implies that black holes have a temperature and an entropy, both of which are proportional to the area of the event horizon \cite{HK+75}. By exploring the thermodynamic relationship between energy, temperature, and entropy, Hawking validated Bekenstein's hypothesis, leading to the precise determination of the proportionality constant that relates a black hole's entropy to the area of its event horizon.

If black holes have entropy, and since black holes are basically a region of spacetime with a certain curvature, then gravity itself could have entropy. Penrose proposed that this property could be quantified by the Weyl curvature tensor, which measures the curvature of spacetime independent of matter \cite{Penrose1979}.

Classical estimators of gravitational entropy (GE) based on curvature tensors have been proposed by several researchers \cite{Rudjord_2008,Romero_2011,cli+13}. For instance, Rudjord, Grøn, and Hervik (RGH) \cite{Rudjord_2008} defined a GE estimator based on the ratio of the Weyl scalar to the Kretschmann scalar, showing consistency with Bekenstein-Hawking entropy \cite{Rudjord_2008}. Romero, Thomas, and Pérez \cite{Romero_2011} applied the RGH proposal to black holes and wormholes, validating its effectiveness in various compact objects, although modifications were necessary for rotating bodies to avoid divergences in the entropy density.

Clifton, Ellis and Tavakol (CET) introduced an alternative estimator of the GE, based on the square-root of the Bel-Robinson tensor, which serves as an effective energy–momentum tensor for free gravitational fields \cite{cli+13}. This measure is non-negative, observer-dependent, and applicable in various cosmological scenarios. It agrees with the Bekenstein-Hawking entropy for Schwarzschild black holes and increases with structural anisotropy in scalar perturbations of a Robertson–Walker universe. When applied to inhomogeneous Lemaître-Tolman-Bondi (LTB) models, it suggests that GE increases with rising inhomogeneities, offering insights into the thermodynamic behavior of gravity across different astrophysical contexts \cite{cli+13}. 

Pérez et al. applied the CET estimator to Kerr black holes, finding its effectiveness depends on the choice of reference frame \cite{Kerr1,Kerr2}. 

Chakraborty et al. \cite{Chakraborty2022} further evaluated the RGH and CET estimators for traversable wormholes. They showed that both provide viable measures of GE: the RGH is particularly effective for rotating systems, while the CET approach is generally effective in many astrophysical and cosmological applications, faces challenges in defining the gravitational temperature for certain geometries, such as the Ellis wormhole.


The goal of this work is to analyze the compatibility of the RGH and CET estimators of GE with the second law of black hole thermodynamics during two conversion processes: the conversion of a Schwarzschild black hole into a Reissner-Nordström black hole by capturing a charged particle, and the reverse process, where the latter becomes a Schwarzschild black hole by absorbing a particle with an electric charge of equal magnitude but opposite sign to the former\footnote{Recently, Quevedo showed \cite{Que24} that the absorption of particles by a rotating black hole increases the irreducible mass of the black hole.}. The article is structured as follows: in section \ref{sec:2} we revisit the concept of irreducible mass, showing that during the transformation from a Schwarzschild black hole to a Reissner-Nordström black hole (and vice versa), the irreducible mass either remains constant (reversible process) or increases (irreversible process). Next, we provide a detailed description of the RGH and CET estimators (section \ref{sec:3}). We apply the CET proposal to estimate the GE of a Reissner-Nordström black hole in section \ref{sec:4}, and  in section \ref{subsec:5a} we evaluate the validity of the second law of thermodynamics during the transformation from a Schwarzschild to a Reissner-Nordström black hole based on predictions from the RGH and CET models. In section \ref{subsec:5b} we address the same problem in the inverse process. The last section is devoted to the discussion of the results obtained and we also summarize the conclusions of the work.

\section{Irreducible mass in black hole transformations}\label{sec:2}

The irreducible mass of a black hole, denoted $M_{\mathrm{irr}}$, is a quantity defined in terms of the mass energy $M$, angular momentum $L$ and charge $Q$ of the black hole as \cite{Ch+70,PhysRevD.4.3552}



\begin{equation}\label{mir}
 M^2 = \left(\ M_{\mathrm{irr}} + \frac{Q^2}{4 \; M_{\mathrm{irr}}}\right)^2 +  \frac{L^2}{4 \; M^2_{\mathrm{irr}}},
\end{equation}

In black hole physics, there are no classical processes that can decrease the irreducible mass of a black hole; therefore, only two black hole transformations are allowed: reversible ones, which do not change the irreducible mass of the black hole, and those that increase it, i.e., irreversible transformations.

Note that the mass energy $M$, charge $Q$, and angular momentum $L$ of a black hole can decrease, but what can never decrease is the corresponding irreducible mass. This is a restatement of the second law of black hole dynamics, since the surface area $A$ of a black hole is 
\begin{equation}
 A = 16 \pi  M_{\mathrm{irr}}^2.
\end{equation}

In the following, we show that in the process of transformation from a Schwarzschild to a Reissner-Nordström black hole, the irreducible mass of the black hole remains unchanged or increases when a charge test particle falls into the former. The same holds for the inverse process. 

We assume that a test particle of rest mass $\mu$, charge $e$ and axial component of the angular momentum $L_z$ is outside a Reissner-Nordström black hole of mass $M$ and charge Q. The energy of the test particle at infinity is computed as (in geometrized units $G= c = 1$) \footnote{The general expression for the energy $E$ of a test particle outside a Kerr-Newman black hole can be found in \cite{Misner1973}}
\begin{equation}\label{energy-rn}
 E = \frac{\beta + \sqrt{\beta^2 - \alpha \gamma}}{\alpha}, 
\end{equation}
where
\begin{eqnarray}
 \alpha & = & r^4,\\
 \beta & = & e Q r^3,\\
 \gamma & = & \left(e Q r\right)^2 - \Delta {L^2}_z - \mu^2 \Delta \rho^2 - \rho^2 \left[(p^r)^2 + (p^{\theta})^2 \right],\\
 \Delta & = & r^2 - 2 M r + Q^2,\\
 \rho^2 & = & r^2,\\
 p^r & =& \frac{dr}{d\lambda},\\
  p^{\theta} & =& \frac{d\theta}{d\lambda}.\\
\end{eqnarray}
We further assume that the particle orbit is in the equatorial plane ($\theta = \pi/2$), so $p^{\theta} = 0$. At the outer event horizon $r = r_+$ ($\Delta =0$), there is a minimum value of $E$, which is
\begin{equation}
 E = \frac{e Q}{r_{+}} +  \lvert p^{r} \rvert.
\end{equation}
Once the particle falls into the black hole, the change in the mass of the black hole is equal to the energy of the particle absorbed. Using $e = \delta Q$
\begin{equation}\label{13}
 \delta M = \frac{Q \delta Q}{r_{+}} +   \lvert p^{r} \rvert.
\end{equation} 
From the positivity of $\lvert p^{r} \rvert$ we deduce that
\begin{equation}
  \delta M \ge  \frac{Q \delta Q}{r_{+}} = \frac{Q\delta Q}{M+\sqrt{M^{2}-Q^{2}}}. 
\end{equation}
Using the latter equation and taking the equality, the resolution of the differential equation leads to expression \eqref{mir}
\begin{equation}
    M=M_{\mathrm{irr}}+\frac{Q^{2}}{4M_{\mathrm{irr}}},
    \label{masa}
\end{equation}
where
\begin{equation}
    M_{\mathrm{irr}}= \frac{1}{2}r_{+}.
\end{equation}

To show that $\delta M_{\mathrm{irr}}\geq 0$, we calculate the differential of $M_{\mathrm{irr}}$ as
\begin{equation}
     2\delta M_{\mathrm{irr}}=\delta M+\frac{M}{\sqrt{M^2-Q^2}}\delta M-\frac{Q}{\sqrt{M^2-Q^2}}\delta Q.
     \label{A01}
\end{equation}
Now, we rewrite Equation \eqref{13} as
\begin{eqnarray}
     r_{+}\delta M-Q\delta Q &= & r_{+}|p^{r}|,\\
     (M+\sqrt{M^{2}-Q^{2}})\delta M-Q\delta Q & = &(M+\sqrt{M^{2}-Q^{2}})|p^{r}|.
\end{eqnarray}
Dividing both sides of the equality by $\sqrt{M^{2}-Q^{2}}$ and simplifying we get
\begin{equation}
    \delta M+\frac{M}{\sqrt{M^2-Q^2}}\delta M-\frac{Q}{\sqrt{M^{2}-Q^{2}}}\delta Q=|p^{r}|\left(1+\frac{M}{\sqrt{M^2-Q^2}}\right).
\end{equation}
By comparing with expression \eqref{A01}, we arrive to
\begin{equation}
 \delta M_{\mathrm{irr}}=\frac{|p^{r}|}{2}\left(1+\frac{M}{\sqrt{M^2-Q^2}}\right)\geq 0.
     \label{117}
\end{equation}

So if we start with a Schwarzschild black hole ($Q = 0$), we see from Eq. \eqref{masa} that $M = M_{\mathrm{irr}}$. When a charge test particle transforms the black hole into a Reissner-Nordström one, according to expression \eqref{117} 
\begin{equation}
     \delta M_{\mathrm{irr}}=|p^{r}|,
     \label{023}
\end{equation}
so,
\begin{equation}
    M_{\mathrm{irrRN}}= |p^{r}| +  M_{\mathrm{irrS}},
    \label{024}
\end{equation}
where $M_{\mathrm{irrRN}}$ and $M_{\mathrm{irrS}}$ are the irreducible masses for Reissner-Nordström and Schwarzschild, respectively, and the corresponding entropy of their event horizons are
\begin{eqnarray}
     S_{\mathrm{S}} & = & k 16\pi M_{\mathrm{irrS}}^{2},\label{025} \\
    S_{\mathrm{RN}} & = & k16\pi M_{\mathrm{irrRN}}^{2}= k16\pi\left(|p^{r}|^{2}+2M_{\mathrm{irrS}}|p^{r}|+M_{\mathrm{irrS}}^{2}\right).\label{026}
\end{eqnarray}
Subtracting equation \eqref{026} from equation \eqref{025} we get
\begin{equation}
S_{\mathrm{RN}}-S_{\mathrm{S}}= k16\pi\left(|p^{r}|^{2}+2M_{\mathrm{irrS}}|p^{r}|\right).
     \label{028}
\end{equation}
Note that this quantity is never negative, which implies that entropy never decreases.

Now we transform this Reissner-Nordström black hole into a Schwarzschild black hole when a test particle of charge $-e$ passes through the event horizon. In this process, the irreducible mass of the black hole increases as follows
\begin{equation}
\delta M_{\mathrm{irr}}=\frac{|p^{r}|}{2} w, \label{035}
\end{equation}
with
\begin{equation}
w=1+\frac{M_{\mathrm{RN}}}{\sqrt{M_{\mathrm{RN}}^2-e^2}}.
\end{equation}
At the end of the process, 
\begin{equation}
    M_{\mathrm{irrS2}}=\frac{|p^{r}|}{2}w+M_{\mathrm{irrRN}},
    \label{036}
\end{equation}
where $M_{\mathrm{irRN}}$ and $M_{\mathrm{irS2}}$ are the irreducible masses for Reissner-Nordström and the new Schwarzschild ($S2$), respectively. The entropy of the event horizon of a Reissner-Nordström black hole is
\begin{equation}
     S_{\mathrm{RN}}=k16\pi M_{\mathrm{irrRN}}^{2},
     \label{037}
\end{equation}
while for the transformed Schwarzschild black hole, denoted $S_{\mathrm{S2}}$ reads
\begin{equation}
     S_{\mathrm{S2}}=k16\pi\left(\frac{|p_{2}^{r}|^{2}}{4}w^{2}+w|p_{2}^{r}|M_{\mathrm{irrRN}}+M_{\mathrm{irrRN}}^{2}\right).
     \label{039}
\end{equation}
Subtracting Eq. \eqref{037} from Eq. \eqref{039} we obtain
\begin{equation}
     S_{\mathrm{S2}}-S_{\mathrm{RN}}=k16\pi\left(\frac{|p_{2}^{r}|^{2}}{4}w^{2}+w|p_{2}^{r}|M_{\mathrm{irrRN}}\right).
     \label{040}
\end{equation}
Again, this quantity is non-negative. Therefore, the entropy never decreases in this inverse transformation process.

In short, we have seen that the entropy or irreducible mass remains the same or increases. How does the mass of the black hole change in these two transformations?

\begin{itemize}
\item First transformation: Schwarzschild to Reissner-Nordström.

The mass energy and irreducible mass of a Reissner-Nordström black hole are related by
\begin{equation}\label{mrn}
M_{\mathrm{RN}} = M_{\mathrm{irrRN}} + \frac{Q^{2}_{\mathrm{RN}}}{4 M_{\mathrm{irrRN}}}.
\end{equation}
At the end of the process, the charge of the Reissner-Nordström black hole corresponds to the charge of the particle, that is, $Q_{\mathrm{RN}} = e$.

The increment of the irreducible mass is given by
\begin{equation}
M_{\mathrm{irrRN}} = |p^{r}_{1}| + M_{\mathrm{irrS1}}.
\end{equation}
Here, $p^{r}_{1}$ is the radial momentum of the particle (as to distinguish from the second particle in the inverse process). Substituting the latter expression into the Eq. \eqref{mrn} yields
\begin{equation}\label{RN1}
M_{\mathrm{RN}} = |p^{r}_{1}| + M_{\mathrm{irrS1}} + \frac{e^2}{4 \left(|p^{r}_{1}| + M_{\mathrm{irrS1}}\right)}.
\end{equation}
If the process is reversible, then $|p^{r}_{1}| = 0$ and $M_{\mathrm{irrRN}} = M_{\mathrm{irrS1}}$. The energy mass of the black hole at the end of the transformation is
\begin{equation}\label{massRN}
M_{\mathrm{RN}} =  M_{\mathrm{irrS1}} + \frac{e^2}{4 M_{\mathrm{irrS1}}}.
\end{equation}
Thus, even in a reversible process $M_{\mathrm{RN}} > M_{\mathrm{S1}} = M_{\mathrm{irrS1}}$.

\item Second transformation: Reissner-Nordström to Schwarzschild.

From Eq. \eqref{036} and since $M_{\mathrm{irrS2}} = M_{S2}$, the energy mass of the Schwarzschild black hole is
\begin{equation}\label{secondt}
M_{\mathrm{S2}}=\frac{|p^{r}_{2}|}{2} \left(1+\frac{M_{\mathrm{RN}}}{\sqrt{M_{\mathrm{RN}}^2-e^2}}\right)+M_{\mathrm{irrRN}}.
\end{equation}
The final mass of the new Schwarzschild black hole depends on whether the transformations are reversible or not. In fact, we can distinguish four different cases: 
\begin{enumerate}
    \item If $|p^{r}_{1}| = 0$, then $M_{\mathrm{S1}} = M_{\mathrm{irrRN}}$, and if also $|p^{r}_{2}| = 0$, $M_{\mathrm{S2}} = M_{\mathrm{irrRN}}$. Then, $M_{\mathrm{S1}} = M_{\mathrm{S2}}$ so both the first and second Schwarzschild black holes are indistinguishable.

    \item If $|p^{r}_{1}| = 0$, then $M_{\mathrm{S1}} = M_{\mathrm{irrRN}}$; but if in the second process  $|p^{r}_{2}| \neq 0$, then from Eq. \eqref{secondt}

 
    \begin{equation}\label{secondt2}
    M_{\mathrm{S2}}=\frac{|p^{r}_{2}|}{2} \left(1+\frac{M_{\mathrm{RN}}}{\sqrt{M_{\mathrm{RN}}^2-e^2}}\right)+M_{\mathrm{S1}},
    \end{equation}
    where $M_{\mathrm{RN}}$ is given by expression \eqref{massRN}. Clearly, $M_{S2} > M_{S1}$. 
    
    \item If $|p^{r}_{1}| \neq 0$ the mass energy of the Reissner-Nordström black hole follows Eq. \eqref{RN1}. Now, if the second transformation is reversible ($|p^{r}_{2}| = 0$), we see from Eq. \eqref{secondt} that
    \begin{equation}
    M_{\mathrm{S2}} = M_{\mathrm{irrRN}} = |p^{r}_{1}| + M_{\mathrm{S1}}.
    \end{equation}
    Again, $M_{S2} > M_{S1}$.
    \item If $|p^{r}_{1}| \neq 0$ and also $|p^{r}_{2}| \neq 0$, then
    \begin{equation}
    M_{\mathrm{S2}}=\frac{|p^{r}_{2}|}{2} \left(1+\frac{M_{\mathrm{RN}}}{\sqrt{M_{\mathrm{RN}}^2-e^2}}\right)+M_{\mathrm{irrRN}},
    \end{equation}
    where
    \begin{eqnarray}
    M_{\mathrm{RN}} & = & |p^{r}_{1}| + M_{\mathrm{S1}} + \frac{e^2}{4 \left(|p^{r}_{1}| + M_{\mathrm{S1}}\right)},\\
    M_{\mathrm{irrRN}} & = & |p^{r}_{1}| + M_{\mathrm{S1}}.
    \end{eqnarray}
\end{enumerate}
As expected, $M_{\mathrm{S2}} > M_{\mathrm{S1}}$.
\end{itemize}

In the next section, we will briefly describe two different proposals for classical estimators of the gravitational entropy.


\section{Classical estimators of gravitational entropy} \label{sec:3}

\subsection{Weyl-Kretschmann estimator}

Rudjord and collaborators \cite{Rudjord_2008} proposed an estimator for the gravitational entropy taking into account the Bekenstein-Hawking entropy. Since the latter is proportional to the area of the event horizon, they suggested that it can be expressed as a surface integral,
\begin{equation}
    S=k_{\mathrm{S}}\int_{\sigma}\Vec{\Psi}\cdot d\Vec{\sigma},
    \label{01}
\end{equation}
where $\Vec{\sigma}$ is the surface area of the event horizon of the black hole, and the vector function $\Vec{\Psi}$ is defined as
\begin{equation}
\Vec{\Psi}=P\hat{e}_{r},
\label{02}
\end{equation}
with $\hat{e}_{r}$ a radial unit vector. The scalar function $P$ is defined using the Weyl and Kretschmann scalars,
\begin{equation}
P^{2}=\frac{C^{abcd}C_{abcd}}{R^{abcd}R_{abcd}}.
\end{equation}

For the proposed description to be viable, it is necessary that on the horizon of a black hole, the surface integral (\ref{01}) reproduces the Bekenstein-Hawking entropy. When Rudjord et al. applied the estimator to the Schwarzschild black hole and, since $P = 1$, they found the value for $k_{\mathrm{S}}$:
\begin{equation}
    k_{\mathrm{S}}=\frac{k_{\mathrm{B}}c^{3}}{4\hbar G},
    \label{04}
\end{equation}
where $k_{\mathrm{B}}$ is the Boltzmann's constant, $G$ is the constant of gravitation and $\hbar = h/2 \pi$ being $h$ the Planck constant.

An expression for the gravitational entropy density can be derived by means of Gauss's divergence theorem
\begin{equation}
s:=k_{\mathrm{S}}|\nabla \cdot \Vec{\Psi}|.
\label{05}
\end{equation}

This estimator was applied to a Reissner-Nordström black hole \cite{Rudjord_2008,Romero_2011}. It was found that the gravitational entropy at the outer horizon is ($G = c = 1$)
\begin{eqnarray}
S_{+} & = & k_{RN}4\pi r^{2}_{+} P, \label{21}\\
P & = & \sqrt{1-\frac{Q^{4}}{6M^{2}r_{+}^{2}-12Mr_{+}Q^{2}+7Q^{4}}}
\label{22}
\end{eqnarray}
where $r_{+}$ is the radius of the outer event horizon and  $M$ and $Q$ are the mass and charge of the black hole. The scalar $P < 1$ so the estimator fails to reproduce the Hawking-Bekenstein entropy at the outer event horizon. In sections \ref{subsec:5a} and \ref{subsec:5b} we will show whether it satisfies the second law in the transformation processes of interest.

\subsection{Bel-Robinson estimator}

The definition proposed by Clifton et al. \cite{cli+13} is motivated by an analogy with classical thermodynamics,
\begin{equation}
T_{\mathrm{grav}}dS_{\mathrm{grav}}=dU_{\mathrm{grav}}+p_{\mathrm{grav}}dV,
\label{06}
\end{equation}
where $T_{\mathrm{grav}}$, $S_{\mathrm{grav}}$, $U_{\mathrm{grav}}$, and $p_{\mathrm{grav}}$ denote the effective temperature, entropy, internal energy, and isotropic pressure of the gravitational field, respectively, and $V$ is the spatial volume.

From the Bel-Robinson tensor, expressions are derived for the effective energy density $\rho_{\mathrm{grav}}$ and for the isotropic pressure $p_{\mathrm{grav}}$, and these depend on whether the spacetime is Coulomb-like (Petrov type D) or wave-like (Petrov type N).

The expressions for $p_{\mathrm{grav}}$ and $\rho_{\mathrm{grav}}$ for Coulomb-like spacetimes (which include black hole spacetimes) have the following form:
\begin{equation}
8\pi \rho_{\mathrm{grav}}=2\alpha\sqrt{\frac{2W}{3}},
\label{07}
\end{equation}
\begin{equation}
p_{\mathrm{grav}}=0,
\label{08}
\end{equation}
with $\alpha$ as a constant, and $W$ is the \textit{super-energy density}
\begin{equation}
W=\frac{1}{4}\left(E_{a}^{b}E_{b}^{a} + H_{a}^{b}H_{b}^{a}\right),
\label{09}
\end{equation}

where $E_{ab}$ and $H_{ab}$ are the electric and magnetic parts of the Weyl tensor, respectively.

To define the temperature, results from black hole thermodynamics and quantum field theory in curved spacetimes are used. The definition is such that it is local (and not only valid for horizons) and reproduces in the correct limits the results from semiclassical calculations in Schwarzschild and de Sitter spacetimes. It has the following expression:
\begin{equation}
T_{\mathrm{grav}}=\frac{|u_{a;b}l^{a}k^{b}|}{\pi}.
\label{010}
\end{equation}
Here,
\begin{equation}
k^{a}=\frac{1}{\sqrt{2}}(u^{a}+z^{a}),
\label{011}
\end{equation}
\begin{equation}
l^{a}=\frac{1}{\sqrt{2}}(u^{a}-z^{a}).
\label{012}
\end{equation}
where $z^{a}$ is a spacelike unit vector and $u^{a}$ is a timelike unit vector aligned with the Weyl principal direction.

In the following section we calculate for the first time the Bel-Robinson estimator for a Reissner-Nordström black hole.


\section{Bel-Robinson  estimator for a Reissner-Nordström black hole}\label{sec:4}

Let us consider the Reissner-Nordström spacetime metric in Gullstrand-Painlevé coordinates $(T, r, \theta, \phi)$:
\begin{equation}
    ds^{2}=-c^{2}\left(1-\frac{2GM}{c^{2}r}+\frac{kGQ^{2}}{c^{4}r^{2}}\right)dT^{2}-2c\left(\sqrt{\frac{2GM}{c^{2}r}-\frac{kGQ^{2}}{c^{4}r^{2}}}\right)drdT+dr^{2}+r^{2}d\Omega^{2},
    \label{0125}
\end{equation}
where $c$, $G$, and $k$ denote the speed of light, the gravitational constant, and Coulomb’s constant, respectively. This coordinate system admits hypersurfaces with constant $T$ that intersect with the event horizon and have Euclidean geometry. For the unit vectors $u^{a}$ and $z^{b}$, the following are chosen:
\begin{equation}
u_{a}=\left(0;\frac{1}{\sqrt{|1-\frac{2GM}{c^{2}r}+\frac{kGQ^{2}}{c^{4}r^{2}}|}},0,0\right),
\label{013}
\end{equation}

\begin{equation}
z^{b}=\left(\frac{1}{c\sqrt{|1-\frac{2GM}{r}+\frac{kGQ^{2}}{c^{4}r^{2}}|}};0,0,0\right).
\label{013}
\end{equation}

We use Eq. (\ref{07}) to find the gravitational energy density. This gives:
\begin{equation}
8\pi \rho_{\mathrm{grav}}=\beta 2\alpha\frac{G|Mc^{2}r-kQ^{2}|}{c^{4}r^{4}}.
\label{014}
\end{equation}

To determine the temperature, we calculate the vectors $l$ and $k$ based on the vectors $u$ and $z$ using expressions (\ref{011}) and (\ref{012}), respectively, and then apply equation (\ref{010}):
\begin{equation}
T_{\mathrm{grav}}=\beta\frac{G|Mc^{2}r-kQ^{2}|}{c^{4}2\pi r^{3}\sqrt{|1-\frac{2GM}{c^{2}r}+\frac{kGQ^{2}}{c^{4}r^{2}}|}},
\label{015}
\end{equation}
where $\beta$ is a factor introduced to ensure that the units of gravitational energy density and gravitational temperature are consistent. It is given by:
\begin{equation}
\beta= \frac{\hbar c}{k_{B}}.
\label{016}
\end{equation}

As explained in \cite{cli+13}, a small variation in gravitational entropy occurs when a small amount of mass is added to the black hole
\begin{equation}
\delta s_{\mathrm{grav}}=\frac{\delta(\rho_{\mathrm{grav}}v)}{T_{\mathrm{grav}}}.
\label{017}
\end{equation}
That is, an increase in effective energy density at a given temperature $T_{\mathrm{grav}}$ necessarily implies an increase in the black hole's gravitational entropy. The integration of Eq. \eqref{017} over the volume $V$ enclosed by the event horizon (where $v=z^{a}\eta_{abcd}dx^{b}dx^{c}dx^{d}$) yields
\begin{equation}
S_{\mathrm{grav}}=\int_{V}s_{\mathrm{grav}}=\int_{V}\frac{\rho_{\mathrm{grav}}v}{T_{\mathrm{grav}}}=\int_{V}\frac{\alpha r}{2}\sin\theta dr d\theta d\phi,
\label{018}
\end{equation}
\begin{equation}
S_{\mathrm{grav}}=\alpha\frac{A_{hor}}{4}.
\label{019}
\end{equation}

Comparing Eq. (\ref{019}) with the expression for Bekenstein-Hawking entropy, we see that:
\begin{equation}
    \alpha=\frac{k_{B}c^{3}}{\hbar G}.
    \label{020}
\end{equation}
Thus, we have shown that the gravitational entropy computed with the Bel-Robinson estimator coincides with the Hawking-Bekenstein entropy for a Reissner-Nordström black hole. In the following, we analyze the behaviour of $\rho_{\mathrm{grav}}$ and $T_{\mathrm{grav}}$. 

We first substitute $\alpha$ and $\beta$ into Eqs. (\ref{014}) and (\ref{015}) 
\begin{eqnarray}
8\pi \rho_{\mathrm{grav}} & = & 2\frac{|Mc^{2}r-kQ^{2}|}{r^{4}}, \label{021}\\
T_{\mathrm{grav}} & = & \frac{\hbar G}{k_{B}c^{3}2\pi r^{3}} \frac{|Mc^{2}r-kQ^{2}|}{\sqrt{|1-\frac{2GM}{c^{2}r}+\frac{kGQ^{2}}{c^{4}r^{2}}|}}. \label{022} 
\end{eqnarray}


In the limit $Q \rightarrow 0$, the corresponding expressions for a Schwarzschild  black hole, as obtained by Clifton and coworkers, are recovered \cite{cli+13}.

Note that in order to have a black hole spacetime geometry, the relationship between mass and charge must satisfy
\begin{equation}
    Q=\lambda M\sqrt{\frac{G}{k}},
    \label{Atom20}
\end{equation}
with $\lambda \leq 1$.

The gravitational energy density and gravitational temperature are zero both when $r$ approaches infinity and when the radius takes the following value
\begin{equation}
r_{0}=\frac{kQ^{2}}{Mc^{2}} = \lambda^{2}\frac{MG}{c^{2}}=\lambda^{2}r_g.
\label{Atom21}
\end{equation}

We now consider a Reissner-Nordström black hole with mass $M=10M_{\odot}$, where ${M_{\odot}}$ denotes a solar mass, and three different values for the charge $Q_{1}=0.1M\sqrt{G /k}$, $Q_{2}=0.5M\sqrt{G / k}$ and $Q_{3}=0.9M\sqrt{G/k}$, respectively; condition (\ref{Atom20}) is satisfied in the three cases considered.

\begin{figure}[H]
    \centering
    \hspace*{-0.7cm}
    \includegraphics[scale=0.495]{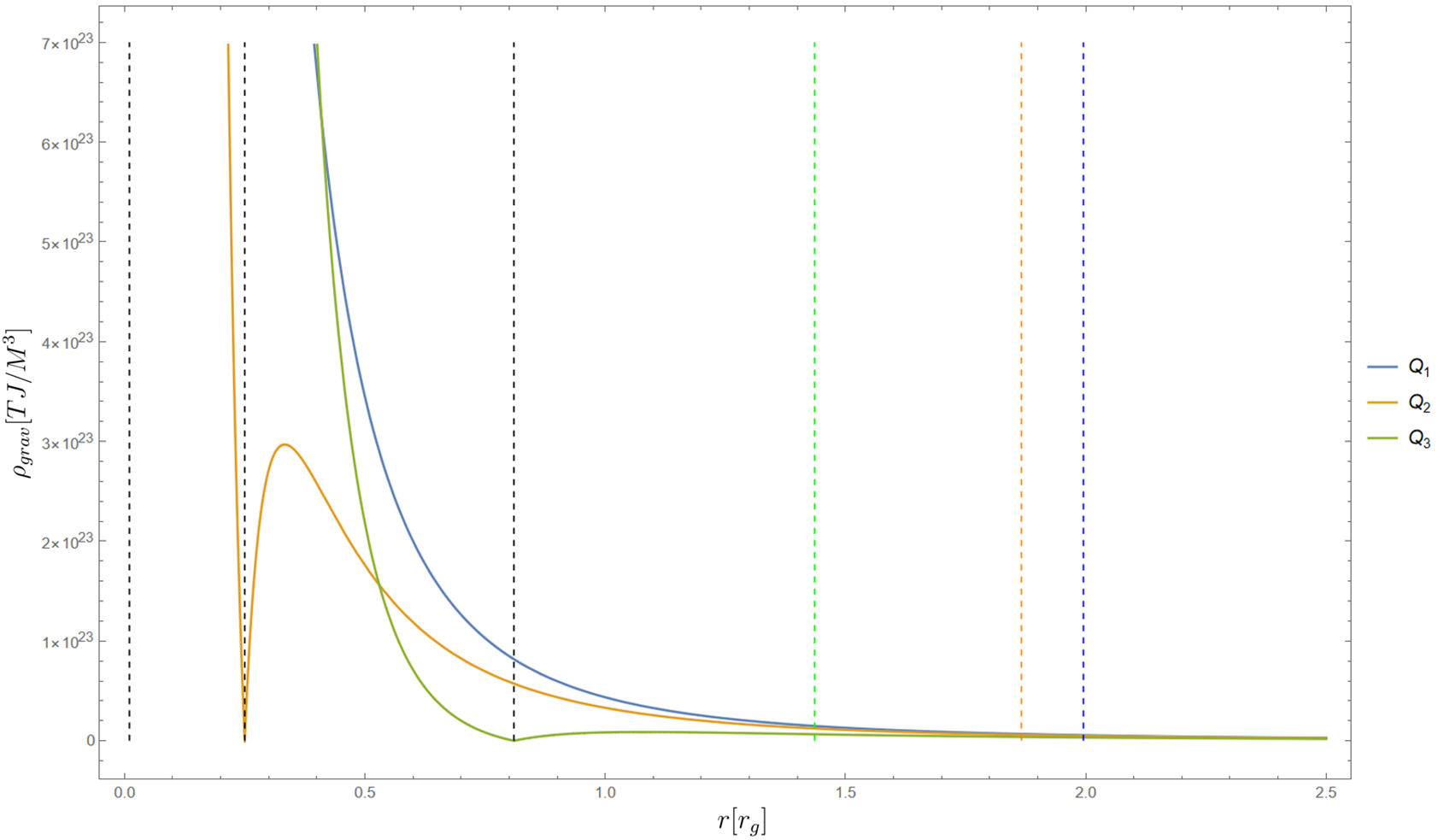}
    \caption{Plot of the gravitational energy density as a function of radius. Here, $M=10M_{\odot}$. The values of the charges are $Q_{1}=0.1M\sqrt{G /k}$, $Q_{2}=0.5M\sqrt{G / k}$, and $Q_{3}=0.9M\sqrt{G/k}$.}
    \label{d02}
\end{figure}

\begin{figure}[H]
    \centering
    \hspace*{-0.5cm}
    \includegraphics[scale=0.47]{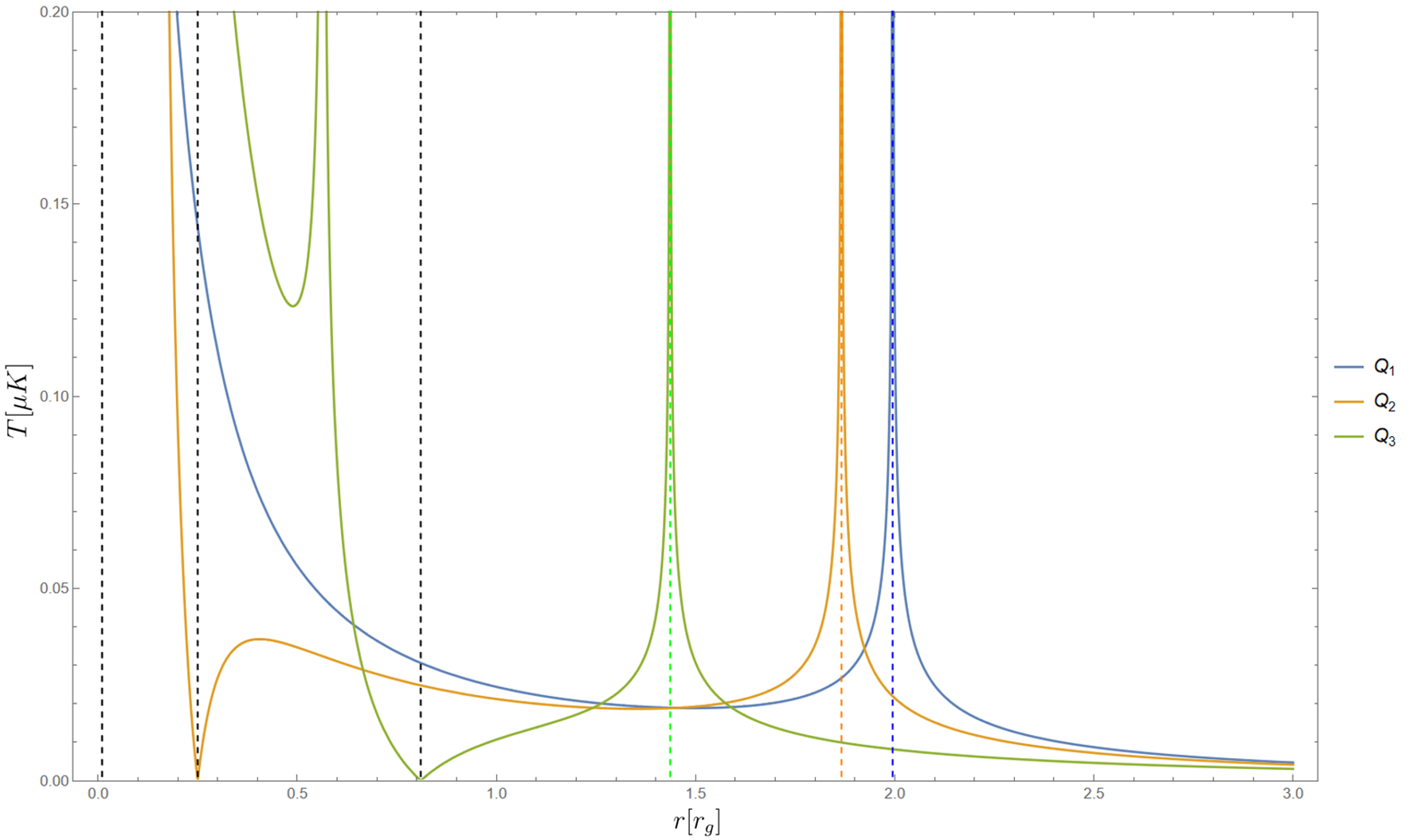}
    \caption{Plot of gravitational temperature as a function of radius. Here, $M=10M_{\odot}$. The values of the charges are $Q_{1}=0.1M\sqrt{G /k}$, $Q_{2}=0.5M\sqrt{G / k}$, and $Q_{3}=0.9M\sqrt{G/k}$.}
    \label{d12}
\end{figure}

We show in Figs. \ref{d02} and \ref{d12} plots of the gravitational energy density and gravitational temperature for three different values of the charge. The black hole with charge $Q_{1}$ is identified with the blue curve, while $Q_{2}$ and $Q_{3}$ with the orange and green curves, respectively. The dotted lines correspond to the external event horizon of the black holes.

From Figure (\ref{d02}), we see that the gravitational density is infinite at the essential singularity ($r = 0$) and finite on both the internal and external event horizons. In the limit $r \rightarrow \infty$, $\rho_{\mathrm{grav}} \rightarrow 0$ as expected.

Note that the gravitational energy density is larger for the black hole with charge $Q_1$ on its event horizon compared to the black holes with charges $Q_2$ and $Q_3$ on their respective event horizons. In other words
\begin{equation}
    \rho_{\mathrm{grav}}(r \geq r_{+1},Q_1) \geq \rho_{\mathrm{grav}}(r \geq r_{+2},Q_2) \geq \rho_{\mathrm{grav}}(r \geq r_{+3},Q_3).
    \label{Atom10}
\end{equation}

In Fig. \ref{d02}, the black dashed lines indicate the radial values ($r_{0}$) where the energy density is zero. In the region inside the outer event horizon, the gravitational energy density reaches a maximum. In the plot, it is only visible for $Q_{2}$, but it is also the case for the other charged black holes. We compute the radial location of the maxima as 
\begin{equation}
\frac{d \rho_{\mathrm{grav}}}{d r}=\frac{3Mc^{2}r-4kQ^{2}}{4\pi r^{5}} = 0,\;\; \Rightarrow \;\; r_{\mathrm{max}}=\frac{4kQ^{2}}{3Mc^{2}}.
\end{equation}




Substituting Eq. \eqref{Atom20} into $r_{\mathrm{max}}$ and simplifying, we get:
\begin{equation}
r_{\mathrm{max}}=\frac{4}{3}\lambda^{2}r_g.
\end{equation}
From this expression, we verify that the maximum occurs after the value where the energy density is zero ($r_{\mathrm{max}}> r_{0} = \lambda^{2}r_g$). Evaluating $r_{\mathrm{max}}$ in expression \eqref{021} and simplifying, we find the maximum value of the gravitational energy density
\begin{equation}
\rho_{\mathrm{max}}=\frac{3^{3}M^{4}c^{8}}{4^{5}\pi k^{3}Q^{6}}.
\end{equation}

It is interesting that this value corresponds to 1/12 of the energy density of a sphere with uniformly distributed energy $Mc^{2}$ and radius $r_{\mathrm{max}}$:

\begin{equation}
\rho_{\mathrm{max}} = \frac{Mc^{2}}{12V_{\mathrm{max}}},
\end{equation}
where
\begin{equation}
V_{\mathrm{max}} = \frac{4\pi r_{\mathrm{max}}^3}{3}.
\end{equation}


From Figure (\ref{d12}), we see that in all cases the temperature is not defined at the origin and on the event horizons (both external and internal). In the limit $r \rightarrow \infty$ the temperature tends to zero.

The black lines denote the values where the gravitational temperature is zero in each black hole. Note that for the green curve, it reaches zero between its internal event horizon and external event horizon. We will now show that this is always the case for a Reissner-Nordström black hole. 

First, we will check that the internal horizon is smaller than the radial value where the temperature is zero ($\lambda^{2}r_g$) inside the black hole.

Since any real positive number less than or equal to 1 increases when its square root is applied, the following holds:
\begin{equation}
   0 \leq 1-\lambda^{2} \leq \sqrt{1-\lambda^{2}} \leq 1.
\end{equation}

Solving for $\lambda^{2}$, we get:
\begin{equation}
   \lambda^{2} \geq 1-\sqrt{1-\lambda^{2}},
\end{equation}
and multiplying both sides by $r_g$:
\begin{equation}
   \lambda^{2}r_g \geq r_g - r_g \sqrt{1-\lambda^{2}} = r_g - \sqrt{r_g^{2} - r_g^{2}\lambda^{2}},\label{Atom30}
\end{equation}


Solving for $\lambda^{2}$ from equation (\ref{Atom20}), we get:
\begin{equation}
   \lambda^{2} = \frac{k}{G} \frac{Q^{2}}{M^{2}}.
\end{equation}

Note that:
\begin{equation}
   r_g^{2}\lambda^{2} = \frac{k}{G} \frac{Q^{2}}{M^{2}} \frac{G^{2}M^{2}}{c^{4}} = \frac{kGQ^{2}}{c^{4}} = \frac{GQ^{2}}{4\pi\epsilon_{0}c^{4}} = q^{2}.
\end{equation}


Substituting $r_g^{2}$ into inequality (\ref{Atom30}), it yields
\begin{equation}
   \lambda^{2}r_g \geq r_g - \sqrt{r_g^{2} - q^{2}} = r_{-}.
\end{equation}

For the external horizon case, it is trivial since $r_+ \geq r_g \geq \lambda^{2}r_g$. Thus, we have shown that the following condition is satisfied
\begin{equation}
  r_- \geq \lambda^{2}r_g \geq r_+.
\end{equation}
We have then proved that the radial coordinate where the gravitational temperature and gravitational energy density are zero (inside the black hole) is between the internal horizon and the external horizon.

Our next task is to analyze whether these two gravitational entropy estimators obey the second law of black hole thermodynamics in the process of transformation from a Schwarzschild to a Reisner-Nordström black hole and vice versa. 
\section{Validity of the second law in black hole transformations using gravitational entropy estimators}\label{sec:5}

\subsection{Transformation of a Schwarzschild to a Reissner-Nordström black hole}\label{subsec:5a}

\subsubsection{Bel-Robinson estimator}

Since the Bel-Robinson estimator reproduces the Hawking-Bekenstein entropy for both a Schwarzschild and a Reissner-Nordström black hole, and we have shown in section \ref{sec:2} that the second law is valid in the black hole transformation process considered here, then the Bel-Robinson estimator correctly describes a system that obeys the second law of black hole thermodynamics in this case.









\subsection{Weyl-Kretschmann Estimator}

The entropy at the event horizon of a Schwarzschild black hole is
\begin{equation}
     S_{\mathrm{S}}=k16\pi M_{\mathrm{irrS}}^{2},
     \label{029}
\end{equation}
while for a Reissner-Nordström black hole with $P<1$ (see Eq. \eqref{21})
\begin{equation}
     S_{\mathrm{RN}}=k16\pi M_{\mathrm{irrRN}}^{2} P =k16\pi \left(|p^{r}|+M_{\mathrm{irS}}\right)^{2} P = k16\pi P \left(|p^{r}|^{2}+2M_{\mathrm{irS}}|p^{r}|+M_{\mathrm{irrS}}^{2}\right).
      \label{031}
\end{equation}

Subtracting equation (\ref{029}) from equation (\ref{031}) yields

 \begin{equation}
     S_{\mathrm{RN}}-S_{\mathrm{S}}= k16\pi\left[(P-1)M_{\mathrm{irrS}}^{2}+ P\left(|p^{r}|^{2}+2M_{\mathrm{irrS}}|p^{r}|\right)\right].
      \label{032}
\end{equation}

For the second law of thermodynamics to be satisfied, the following must hold:
\begin{equation}
     (P-1)M_{\mathrm{irrS}}^{2}+ P\left(|p^{r}|^{2}+2M_{\mathrm{irrS}}|p^{r}|\right) \geq 0,
      \label{033}
\end{equation}

Solving the inequality for $|p^{r}|$ gives
\begin{equation}
   |p^{r}|\geq M_{\mathrm{irrS}}\left[\frac{1}{\sqrt{P}}-1\right].
   \label{034}
\end{equation} 

We see that only for an irreversible processes does the Weyl-Kretschmann estimator ensure that the variation of the gravitational entropy during the transformation process is positive.
\subsection{Transformation of a Reissner-Nordström to a Schwarzschild black hole}\label{subsec:5b}

\subsubsection{Bel-Robinson Estimator}

Here, the same argument provided in section \ref{subsec:5a} holds for the inverse process, so the estimator again represents a physical system that obeys the second law of black hole thermodynamics.

\subsubsection{Weyl-Kretschmann Estimator}
The entropy at the event horizon of a Reissner-Nordström black hole ($P<1$) is
\begin{equation}
     S_{\mathrm{RN}}=k16\pi M_{\mathrm{irrRN}}^{2} P. 
     \label{041}
\end{equation}

After the transformation, the final entropy at the event horizon of the Schwarzschild black hole ($P=1$) yields
\begin{equation}
     S_{\mathrm{S2}}=k16\pi M_{irS2}^{2}=k16\pi \left(\frac{|p_{2}^{r}|}{2}w+M_{\mathrm{irrRN}}\right)^{2} = k16\pi\left(\frac{|p_{2}^{r}|^{2}}{4}w^{2}+w|p_{2}^{r}|M_{\mathrm{irrRN}}+M_{\mathrm{irrRN}}^{2}\right).
     \label{042}
\end{equation}

Subtracting Eq. \eqref{041} from Eq. \eqref{042}, we obtain:
\begin{equation}
     S_{\mathrm{S2}}-S_{\mathrm{RN}}=k16\pi M_{\mathrm{irrRN}}^{2}\left(1-P\right)+k16\pi\left(\frac{|p_{2}^{r}|^{2}}{4}w^{2}+w|p_{2}^{r}|M_{\mathrm{irrRN}}\right).
\end{equation}
In both a reversible and a non-reversible process, there is an increase in gravitational entropy, so that for this particular black hole transformation the Weyl-Kretschmann estimator adequately describes the second law of black hole thermodynamics.
\section{Discussion and conclusions}\label{sec:6}

The goal of this work was to determine whether the second law of black hole thermodynamics is satisfied in certain black hole transformation processes using two specific estimator for the gravitational entropy. First, it was explicitly shown that the entropy of the event horizon remains constant or increases both in the process of transformation from a Schwarzschild black hole to a Reissner-Nordström black hole when a test particle of charge $e$ (in the inverse transformation the charge is $-e$) enters the black hole.

The two estimators analyzed were the Weyl-Kretschmann and the Bel-Robinson. In particular, we showed that the latter reproduces the Hawking-Bekenstein entropy for a Reissner-Nordström black hole. Consequently, this estimator satisfies the second law of black hole thermodynamics in both transformation processes.

It was previously shown \cite{Rudjord_2008,Romero_2011} that while the Weyl-Kretschmann estimator reproduces the Hawking-Bekenstein entropy for a Schwarzschild black hole, it does not in the charged case. The limitations of the Weyl-Kretschmann estimator in the study of Reissner-Nordström black holes could stem from the fact that the Kretschmann scalar incorporates information from the components of the Ricci tensor, which, through Einstein's field equations, is related to the distribution of matter and energy represented by the energy-momentum tensor. In this context, the scalar function P, defined within this estimator, could include information about the gravitational curvature that is directly associated with the local matter and energy. A new definition of the scalar $P$ could help to solve the problem.

Besides these issues, we show that for the first transformation, the second law holds only if the process is irreversible. However, when a second transformation is applied and the black hole becomes uncharged, the estimator obeys the second law for both a reversible and a non-reversible process.




As future work, we will test whether the Bel-Robinson estimator satisfies the second law of thermodynamics in transformation processes from rotating black holes (Kerr spacetime) to rotating black holes with charge (Kerr-Newman spacetime). On the other hand, it would be interesting to analyze the gravitational entropic evolution in cosmological context involving black holes. There are a large number of processes that deserve to be studied and that could lead to results that may allow us to understand a little more about the nature of gravitation and spacetime.

\backmatter

\bmhead{Acknowledgments} G.E.R. thanks Santiago E. Pérez-Bergliaffa for many discussions on black holes. 

A.A.P.L. is deeply grateful with Servando Rojas Garza and Jordán Santillán for the stimulating conversations they regularly have about theoretical physics.

\section*{Declarations}


\begin{itemize}
\item Funding

D. P. acknowledges the support from CONICET under Grant No. PIP 0554 and AGENCIA I$+$D+$i$ under Grant PICT-2021-I-INVI-00387.  A. A. P. L. acknowledges funding from the Consejo Nacional de Humanidades, Ciencias y Tecnologías (CONAHCYT) grant under its program "Becas Nacionales para Estudios de Posgrado". C. L. acknowledges funding from CONAHCYT under its program "Sistema Nacional de Investigadoras e Investigadores (SNII), Mexico".

\item Competing interests 

The authors have no competing interests to declare that are relevant to the content of this article.

\item Ethics approval 

Not applicable

\item Consent to participate

Not applicable

\item Consent for publication

Not applicable

\item Availability of data and materials

Not applicable

\item Code availability 

Not applicable

\item Authors' contributions

All authors contributed equally to the manuscript.

\end{itemize}







\bibliography{MiBiblio}

\end{document}